\documentstyle[aps,floats,prl,preprint]{revtex}

\draft
\tighten     
\begin{document}

\title{Effective Critical Exponents from Finite 
			      Temperature Renormalization Group}
\author{M.~Strickland and S.-B.~Liao}
\address{Department of Physics,
	 Duke University, Durham, NC 27708-0305}

\maketitle

\begin{abstract}
Effective critical exponents for the $\lambda \phi^4$ 
scalar field theory are calculated
as a function of the renormalization group block size $k_o^{-1}$ and 
inverse critical temperature $\beta_c$.  
Exact renormalization group equations are presented
up to first order in the derivative expansion and numerical solutions 
are obtained with and without polynomial expansion of the blocked 
potential.  For a finite temperature system in $d$ dimensions, it is 
shown that $\bar\beta_c = \beta_c k_o$ determines whether the $d$-dimensional 
($\bar\beta_c \ll 1$) or $(d+1)$-dimensional ($\bar\beta_c \gg 1$) fixed 
point governs the phase transition.  The validity of a polynomial 
expansion of the blocked potential near criticality is also addressed.
\end{abstract}

\vspace{2mm}
\narrowtext

The calculation of critical exponents using renormalization group (RG)
methods has a long standing history.  For the scalar $\lambda \phi^4$ 
theory in four dimensions it has been shown that only a trivial Gaussian
fixed point exists, whereas for $2\!\leq\!d\!<4\!$ 
a strong-coupling Wilson-Fisher
fixed point appears \cite{zinn}.  Studies of the transition from 
four- to three-dimensional systems typically rely on the $\epsilon$
expansion \cite{wilson}; however, through years of labor it has been shown
that this expansion is asymptotic, requiring complicated resummations
or outright truncation of the series.  In the last few years a new
tool has emerged for studying phase transitions in arbitrary dimension:
the numerical and/or analytical solution of exact renormalization group 
equations \cite{rg,eijck,morris,attan}.  When solving RG equations
one conventionally Taylor expands the blocked potential and solves for the
flow of the Taylor coefficients;  however, we find evidence
that Taylor expansion of the blocked potential near criticality is not
justified.  In this Letter, we will present results obtained \cite{ourpaper} 
solving finite temperature RG  equations numerically without polynomial 
expansion.

For generality we first consider a finite temperature system 
defined on the manifold $S^1\times R^d$ with the radius of $S^1$ equal to 
the inverse temperature $\beta$. As $T \to 0$, the radius of $S^1$
goes to infinity making the manifold equivalent to $R^{d+1}$,
while in the opposite limit, $T \to \infty$, the manifold becomes
equivalent to $R^{d}$.  However, as we will show below, the critical
behavior of a finite temperature system is actually parameterized by
$\bar\beta_c = \beta_c k_o$, where $k_o$ is the 
renormalization blocking scale.  The critical behavior of the 
system will be characteristic of $(d+1)$-dimensions when $\bar\beta_c \gg
1$ and $d$-dimensions when $\bar\beta_c \ll 1$, so that in the true 
thermodynamic limit, $k_o\!\to\!0$, the system will exhibit $d$-dimensional 
critical behavior for all critical temperatures $T_c > 0$.

The bare action for a scalar field theory is given by
\begin{equation}
S[\phi] = \int_0^{\beta}\!\!\!d\tau\!\!\int\!\!d^d {\bf x} 
\left\{\!{Z_{\tau}\over 2}(\partial_{\tau}\phi)^2 + 
{Z_s \over 2}(\nabla\phi)^2 + V(\phi) \!\right\},
\label{bareaction}
\end{equation}
where we have set $\hbar=1$, and 
$Z_{\tau}$ and $Z_s$ are the tree-level temporal and spatial
wavefunction renormalization constants, respectively.  Note that two
wavefunction renormalization constants are required at finite temperature 
due to the breaking of Lorentz invariance. 

The RG evolution 
of the finite temperature blocked action $\tilde{S}_{\beta,k}[\Phi]$ can
be obtained from
\begin{equation}
{\rm e}^{-\tilde{S}_{\beta,k}[\Phi]} = \int\!\!D[\phi]\prod_{\bf x}
\delta(\phi_{k,\tilde{n}}({\bf x},\tau) - 
\Phi({\bf x},\tau))\,{\rm e}^{-S[\phi]}\,,
\label{path}
\end{equation}
where $\phi_{k,\tilde{n}}({\bf x},\tau)$ is the blocked quantum field,
which in general can be expressed in momentum space as
\begin{equation}
\phi_{k,\tilde{n}}({\bf p},\omega_n)=
\rho_{k,\tilde{n}}^{(d)}({\bf p},\omega_n)\,\phi({\bf p},\omega_n)\,,
\end{equation}
with $\rho_{k,\tilde{n}}^{(d)}({\bf p},\omega_n)$ being the blocking function
for the system.  We choose the blocking function to be
\begin{equation}
\rho_{k,\tilde{n}}^{(d)}({\bf p},\omega_n) = 
\delta_{n=0}\,\Theta(k - \mid\!{\bf p}\!\!\mid)\,,
\end{equation}
so that only the modes with $\left| {\bf p} \right| > k$ and $n=0$ are
included in the blocked field with the other modes being integrated
over in the path integral.    
As for the spatial component of the blocking function, 
a sharp cutoff cleanly separates the high- and low-momentum modes and does
not introduce any dependence on the shape of the blocking function, as is the
case for a smooth cutoff.

In order to derive the coupled RG equations for the blocked potential 
and wavefunction renormalization we evaluate
(\ref{path}) at one-loop level and expand the background field as 
$\Phi({\bf x}) = \Phi_o + \tilde{\Phi}({\bf x})$.  Derivative terms 
are handled by treating the fluctuation field, $\tilde{\Phi}({\bf x})$, and 
the momentum, ${\bf p}$, as operators obeying the commutation 
relation $\left[ p_i , \tilde{\Phi}({\bf x}) \right]  =  
i \partial_i \tilde{\Phi}({\bf x}) $ \cite{fraser}.

We note here that since $\Phi({\bf x})$ is independent of $\tau$ we can 
neglect the RG flow of $Z_\tau$ and make the notational identification 
$Z \equiv Z_s$ throughout the rest of the Letter.
In terms of the bare potential $V(\Phi)$ and bare
wavefunction renormalization $Z$ the one-loop contributions to the 
blocked potential and wavefunction renormalization are
\begin{eqnarray}
\tilde{U}_{\beta,k}^{(1)} & = & {1 \over \beta} 
\int_{\mid{\bf p}\mid=k}^{\Lambda} \ln \left[ {\rm sinh} \left(
{\beta u_{{\bf p},\Phi} \over 2} \right) \right] \nonumber \\
\tilde{{\cal Z}}_{\beta,k}^{(1)} & = & {Z (V^{'''})^2 \over 2 \beta }
\sum_{n=-\infty}^{\infty} \int_{\mid{\bf p}\mid=k}^{\Lambda}
{ - Z {\bf p}^2/3 + \omega_n + V^{''} \over (\omega_n^2 + 
u_{{\bf p},\Phi}^2)^4 }  \, ,
\label{ima}
\end{eqnarray}
where $u_{{\bf p},\Phi} \equiv ( Z\,{\bf p}^2 + V^{''}(\Phi) )^{1/2}$, 
and $\Lambda$ is the ultraviolet cutoff and a tilde indicates a
perturbative quantity.

\begin{figure}[t]
\vspace{2.75in}
\includegraphics{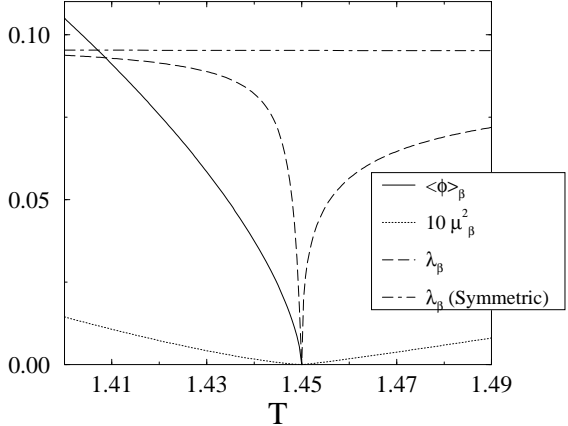}
\vspace{-8mm}
\caption{Temperature dependence of the thermal mass parameter
$\mu^2_{\beta}$, coupling constant $\lambda_{\beta}$, and 
the expectation value $\left<\phi\right>_{\beta}$ near $T=T_c$. The $T$
dependence of $\lambda_{\beta}$ in the symmetric phase is also 
illustrated.}
\label{critfig}
\end{figure}

Due to their perturbative nature, these equations will fail to capture
the true physics near criticality.  This is because as the infrared (IR)
cutoff, $k$, is lowered, the loop expansion of the path integral will no 
longer be expanding about the true 
minimum of the blocked action but instead 
about the minimum of the bare action. As an improvement,
we can take advantage of the arbitrariness of the IR cutoff and 
divide the integration into a large number of thin shells.  For instance,  
we could integrate eq.~(\ref{ima}) from the UV cutoff, 
$\Lambda$, down to $\Lambda - \Delta k$ and use the potential obtained,
$\tilde{U}^{(1)}_{\beta,\Lambda - \Delta k}$, as the initial condition for 
the integration of the next shell.  By dividing the integation into 
thin shells and taking the limit as the shell width $\Delta k$ goes to zero, 
eq.~(\ref{ima}) can be  made self-consistent 
allowing a non-perturbative solution 
of the path integral.  This procedure is equivalent to replacing the bare
potential, $V$, and bare wavefunction renormalization, $Z$, with their $k$
dependent counterparts $U_{\beta,k}$ and ${\cal Z}_{\beta,k}$ on the right
hand side of (\ref{ima}), giving
\begin{eqnarray}
U_{\beta,k}^{(1)} & = & {1 \over \beta} 
\int_{\mid{\bf p}\mid=k}^{\Lambda} \ln \left[ {\rm sinh} \left(
{\beta u_{{\bf p},\Phi} \over 2} \right) \right] \nonumber \\
{\cal Z}_{\beta,k}^{(1)} & = & {{\cal Z}_{\beta,k} (U_{\beta,k}^{'''})^2 \over 
2 \beta }\!\!\!\!\sum_{n=-\infty}^{\infty} 
\int_{\mid{\bf p}\mid=k}^{\Lambda}\!\!\!\!\!{ - 
{\cal Z}_{\beta,k}{\bf p}^2/3 + \omega_n + 
U_{\beta,k}^{''} \over (\omega_n^2 + 
u_{{\bf p},\Phi}^2)^4 }  \, , \nonumber \\
\label{rgint}
\end{eqnarray}
with $u_{{\bf p},\Phi}$ now given by $u_{{\bf p},\Phi} = 
( {\cal Z}_{\beta,k}\,{\bf p}^2 + U_{\beta,k}^{''}(\Phi) )^{1/2}$.  
These equations are exact to all loop orders since higher-loop 
contributions are suppressed by additional powers of $\Delta k$, 
allowing us to sum over all non-overlapping graphs including daisy and super-daisy diagrams \cite{oldpaper}.

\begin{figure}[t]
\vspace{2.75in}
\includegraphics{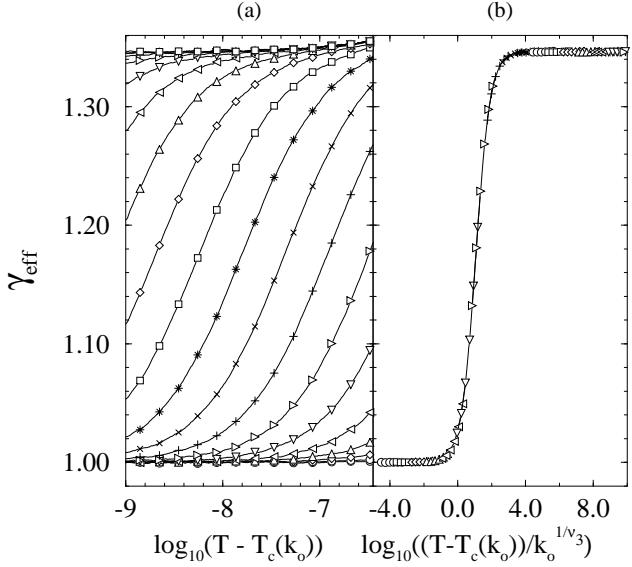}
\vspace{2mm}
\caption{Effective critical exponent 
$\gamma_{\rm eff}$
as function of (a) ${\rm log}_{10}(T - T_c(k_o))$  and (b) 
${\rm log}_{10}((T -$ $T_c(k_o))/k_o^{1/\nu_3})$ at $\Phi^8$ level of
truncation. The different symbols correspond to different values of 
$k_o$.}
\label{scalefig}
\end{figure}

Since eq. (\ref{rgint}) contains two scales,
$k$ and $\beta$, there are two ways to proceed at this stage.  The 
first would be to fix $k$ and differentiate (\ref{rgint})
with respect to the temperature;~
we call this the temperature RG \cite{rg}.  We will not discuss this method 
here but instead will concentrate on the other scale, $k$, which produces
the momentum RG.  After performing the Matsubara sums 
\cite{ourpaper} in
(\ref{rgint}) and differentiating the coupled integral equations
with respect to the $k$ we have

\widetext

\begin{eqnarray}
k {dU_{\beta,k} \over d k}
 & = & - { S_d k^d \over \beta } \ln \left[ {\rm sinh} 
\left( {\beta u_{{\bf p},\Phi} \over 2} \right) \right] \nonumber \\
k {d {\cal Z}_{\beta,k} \over  d k } & = & - {{\cal Z}_{\beta,k} 
(U_{\beta,k}^{'''})^2 S_d k^d 
\over 48 u_{k,\Phi}^7} \left\{ (-{\cal Z} k^2 + 9  U_{\beta,k}^{''}) \left[ 
{1 \over 2} n_{b,k} (1 + \beta u_{k,\Phi}(1 + n_{b,k})) \right]
+ \right. \nonumber \\ 
& & \hspace{7mm} \left. n_{b,k}(1 + n_{b,k})\beta^2 u_{k,\Phi}^2 \left[ (-{\cal Z} k^2 + 
3  U_{\beta,k}^{''})(1 + 2 n_{b,k}) - {2 \over 3} {\cal Z} k^2 \beta
u_{k,\Phi}(1 + 6 n_{b,k} + 6 n_b{,k}^2) \right]
\right\} \, ,
\label{rgpde}
\end{eqnarray}

\narrowtext

\noindent
where $n_{b,k} = ({\rm e}^{\beta u_{k,\Phi}} - 1)^{-1}$
and $S_d = 2/(4 \pi)^{d/2} \Gamma(d/2)$.
With the initial condition $U_{\beta,\Lambda}(\Phi) = V(\Phi)$ these 
equations allow us to integrate down from $k=\Lambda$ to $k=0$ for a fixed 
external temperature keeping track of not only the relevant but the 
irrelevant coupling constants as well.

Typically, the first step made in solving equations like (\ref{rgpde})
is to expand $U_{\beta,k}$ as
\begin{equation}
U_{\beta,k}(\Phi) = \sum_{m=1}^{\infty} { g_{\beta,k}^{(2m)} \over (2m)! } 
\Phi^{2m},\hspace{5mm} g_{\beta,k}^{(2m)} = U_{\beta,k}^{(2m)}(0) \, ,
\end{equation}
and solve the coupled ordinary differential equations
for the coefficients $g_{\beta,k}^{(2m)}$ after a truncation of the series. 
However, it is also possible to solve the coupled equations (\ref{rgpde})
without making any polynomial expansion or truncation.  To accomplish this we 
simply discretize $U_{\beta,k}$ and 
${\cal Z}_{\beta,k}$ in $\Phi$ and $k$ and use finite differences to advance
the blocked potential from the bare form to the renormalized form.  It is
desirable to solve (\ref{rgpde}) without polynomial 
truncation since in general the potential will be a 
non-analytic function of $\Phi$ at the critical point \cite{zinn,eijck} 
as well as to provide a consistency check of the polynomial expansion.

\begin{table}
\begin{tabular}{c|l}
$g_{\beta,k}^{(2m)} \sim k^{a_{m}}$ & $a_m$ \\ 
\hline
$g_{\beta,k}^{(2)}$ & 1.998 $\pm$ 0.005 = $\gamma_{\rm eff}/\nu_{\rm eff}$ \\
\hline
$g_{\beta,k}^{(4)}$ & 0.997 $\pm$ 0.004 = $\zeta_{\rm eff}/\nu_{\rm eff}$ \\
\hline
$g_{\beta,k}^{(6)}$ & 0.002 $\pm$ 0.004  \\
\hline
$g_{\beta,k}^{(8)}$ & -1.001 $\pm$ 0.005  \\
\hline
$g_{\beta,k}^{(10)}$ & -2.002 $\pm$ 0.005  \\
\hline
$<\!\Phi\!>$ & 0.51 $\pm$ 0.01 = $\beta_{\rm eff}/\nu_{\rm eff}$ \\
\end{tabular}
\caption{$k$ dependence of vertex functions at the critical point neglecting
the effects of wavefunction renormalization.}
\label{fstable}
\end{table}

In Figure \ref{critfig} we have plotted the temperature dependence of the
renormalized mass, coupling constant, and expectation value, $\hat{\Phi}$,
for $d=3$. 
The Figure shows that all three of these quantities vanish at $T_c$ 
continuously indicating a second-order phase transition.  Exactly how these
quantities vanish determines the critical exponents.  Since we have
control over both the temperature and $k$, we can
measure the critical exponents in two ways.  The first method is to integrate 
eq. (\ref{rgpde}) down to $k=k_o$ and measure the temperature 
dependence of the thermal parameters near $T_c(k_o)$ which is defined by
the condition 
$\mu_{\beta_c(k_o),k_o}^2\!=\!\hat{\Phi}_{\beta_c(k_o),k_o}\!=\!0$.  
Then using the definitions
\begin{eqnarray}
& & \chi^{-1}  = \mu_{\beta,k_o}^2 \sim |T - T_c(k_o)|^{\gamma_{\rm eff}} \, ,
\hspace{5mm} \beta \to \beta_c(k_o) \, , \nonumber \\
& & \eta_{\rm eff}  =  - { \partial {\cal Z}_{\beta_c,k_o}(
\hat{\Phi}_{\beta_c,k_o}) \over \partial {\rm ln} k_o } \, , 
\hspace{14mm} \beta = \beta_c(k_o) \, , \nonumber \\
& & \hat{\Phi}_{\beta,k_o}  \sim  |T - T_c(k_o)|^{\beta_{\rm eff}} \, ,
\hspace{15mm} \beta \to \beta_c(k_o) \, , \nonumber \\
& & \hat{\Phi}_{\beta_c(k_o),k_o}  \sim  h^{1 / \delta_{\rm eff}} \, , 
\hspace{22mm} h \to 0 \, ,
\label{critdef}
\end{eqnarray}
we can determine the critical exponents.  The exponents measured in this 
way will be necessarily $k_o$ dependent, with the true $d$-dimensional
critical exponents given in the limit $k_o\!\to\!0$.  For $k_o\!\neq\!0$ the
critical exponents measured interpolate between the three- and
four-dimensional values.

Another way to determine the critical exponents is to fix $T$ at the true
critical temperature $T_c=T_c(k_o=0)$ and use the $k_o$ dependence of
the parameters and finite-size scaling laws.
Finite size scaling states that near $T_c$ the scaling variable is $Y =
\ell/\xi$, where $\xi$ is the correlation length of the system and $\ell$
is the size of the finite dimension.
In our analysis there are two finite length scales, the blocking length, 
$k_o^{-1}$, 
and the inverse critical temperature, $\beta_c$.  The scaling variables
associated with these two scales are 
$Y_\beta = |T - T_c(k_o)|/k_o^{1/\nu_{\rm eff}}$ and 
$Y_k  = (\beta_c k_o)^{1/\nu_{\rm eff}} = \bar\beta_c^{1/\nu_{\rm eff}}$, 
respectively, so that in addition to
the standard scaling variable $Y_\beta$ we get finite size scaling also in the
imaginary time direction governed by the dimensionless quantity $\bar\beta_c$.

Using finite-size scaling we fix $T=T_c$ in eq. (\ref{critdef}) and 
replace $|T_c - T_c(k_o)|$ with $k_o^{1/\nu_{\rm eff}}$ giving
\begin{equation}
\chi^{-1}=\mu_{\beta_c,k_o}^2 \sim k_o^{\gamma_{\rm eff}/\nu_{\rm eff} }  \, , 
\hspace{3mm}
\hat{\Phi}_{\beta_c,k_o}  \sim  k_o^{\beta_{\rm eff}/\nu_{\rm eff}} \, .
\label{critdef2}
\end{equation}
After extracting $\nu_{\rm eff}$ by measuring the $k_o$ dependence of 
$T_c(k_o)$ $( \lim_{k_o \to 0} |T_c(k_o) - T_c| 
\sim k_o^{1 / \nu_{\rm eff}})$ we
can determine the effective critical exponents $\gamma_{\rm eff}$ and 
$\beta_{\rm eff}$ from finite size data.  The results obtained for the
finite size scaling exponents are listed in Table \ref{fstable}.  Together
with a measurement of $\nu_{\rm eff}$ these exponents enable us to calculate
$\gamma_{\rm eff}$, $\beta_{\rm eff}$, and $\zeta_{\rm eff}$.

\begin{figure}[t]
\vspace{2.75in}
\includegraphics{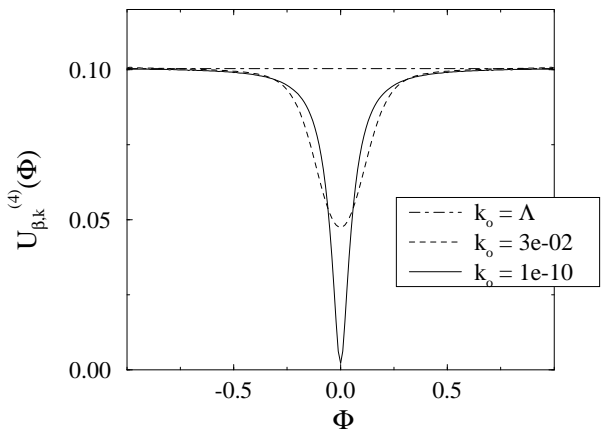}
\vspace{-6mm}
\caption{$U^{(4)}_{\beta,k}$ as a function of $\Phi$ close to $T_c$ 
($T - T_c(k_o)$ $\sim 10^{-5}$) obtained by solving the RG equations
without polynomial truncation.}
\label{polyfig}
\end{figure}

In Table \ref{exptable} we list the critical exponents we calculated along
with previous calculations and experimental values.  The values obtained
using a polynomial truncation of the blocked potential agree with those
obtained in previous works \cite{morris,attan}; however, the
poor agreement with experiment suggests that a 
polynomial truncation of the blocked potential is not appropriate.
This problem has been pointed out by Morris as well \cite{morris}.  
We also observe that the $(2m)$-point functions with $m \geq 3$ diverge with alternating signs 
$g_{\beta_c,k}^{(2m)} \sim (-1)^{(m + 1)} k^{m - 3(1+\eta_{\rm eff})}$.
This pattern of divergence has been reported elsewhere \cite{eijck}
and provides another indication of a non-polynomial structure for the
critical effective potential.

As further evidence for this non-analyticity we can examine the critical
effective potential obtained from our non-truncated code.  In 
Fig.~\ref{polyfig} we have plotted the fourth derivative of the blocked
potential, $U_{\beta,k}^{(4)}$, 
near $T_c$ for various values of $k_o$.  From this Figure we
see that $U_{\beta_c,k\!\to\!0}^{(4)}$ is not identically zero for all $\Phi$
as would be predicted by a fourth order polynomial truncation, i.e.
$U_{\beta_c,k} = \mu_{\beta_c,k}^2 \Phi^2/2 + \lambda_{\beta_c,k} 
\Phi^4/4!$.  Instead, higher order polynomial terms must be
introduced in order to describe the shape of $U_{\beta_c,k\!\to\!0}^{(4)}$.  
For example, a sixth order term would give $U_{\beta_c,k\!\to\!0}^{(4)}$
a parabolic shape near the origin, and higher order terms with increasing
magnitude and oscillating signs are needed to fit the plateau at large $\Phi$.

\widetext 

\begin{table}
\caption{Critical exponents as function of the level of truncation along with
the best calculations to date and experimental values.  In the 
four-dimensional limit mean field exponents are obtained at all levels of
truncation.  NT indicates results
obtained without polynomial truncation of the blocked potential.  Calculations
and experimental values are taken from \protect\cite{zinn} and 
\protect\cite{goldenfeld}, respectively.
}
\vspace{2mm}
\begin{tabular}{|c|c|c|c|c|c|}
\multicolumn{6}{c}{\bf Measured Critical Exponents for $d_{\rm eff} = 3$} \\
\hline
\, $O(\Phi^{(2m)})$ \,& \, $\gamma$ \,&\, $\zeta$ ($\nu$) \,&\, $\beta$ \,
&\, $\eta$ \,&\, $\delta$\, \\
\hline
2 & 1.054 $\pm$ 0.002 & 0.527 $\pm$ 0.003 & & & \\ \hline
3 & 1.171 $\pm$ 0.004 & 0.585 $\pm$ 0.003 & & & \\ \hline
4 & 1.345 $\pm$ 0.004 & 0.672 $\pm$ 0.004 & & & \\ \hline
5 & 1.504 $\pm$ 0.004 & 0.744 $\pm$ 0.004 & & & \\ \hline
NT& 1.234  $\pm$ 0.01 & & 0.315 $\pm$ 0.007 & 0.0000 & 4.65 $\pm$ 0.1 \\ \hline
NT + ${\cal Z}$ & 1.241  $\pm$ 0.008	& & 0.321 $\pm$ 0.008 & 0.036 
     $\pm$ 0.005  & 4.63 $\pm$ 0.1 \\ \hline
\multicolumn{6}{c}{\bf Best Calculations to Date}
\\ \hline
$\epsilon^5$ & 1.2390 $\pm$ 0.0025 & 0.6310 $\pm$ 0.0015 & 0.3270 $\pm$ 
0.0025 & 0.0375 $\pm$ 0.0025 &  4.814 $\pm$ 0.015 \\ \hline
\multicolumn{6}{c}{\bf Experimental Values} \\
\hline
 &  1.23-1.25 & 0.624-0.626 & 0.316-0.327 & 0.016-0.06 & 4.6-4.9 \\ \hline
\end{tabular}
\label{exptable}
\end{table}

\narrowtext

We have demonstrated that it is possible to determine the critical exponents
of the scalar $\lambda \phi^4$ theory
through numerical integration of RG flow equations without polynomial 
expansion.
This method provides a straightforward alternative to the $\epsilon$ expansion and 
can be used to study field theories in arbitrary dimensions.  We have shown
through finite-size scaling arguments that the effective critical exponents
interpolate smoothly between their $(d+1)-$ and $d-$dimensional values.
Our results indicate that the effective potential is a non-analytic function
of the background field at the critical point, 
casting serious doubt on the validity of a polynomial
expansion of the blocked potential in the vicinity of the phase transition.
In future work we will calculate
O(N) exponents and perform studies in two dimensions in order to 
compare our numerical solutions with exact analytic solutions.

\acknowledgements

We thank J.~Adler, R.~Brown, S.~Huang, S.~Matinyan, B.~M\"uller, 
D.~O'Connor and J.~Socolar for stimulating discussions and helpful 
comments. This work was supported in part by the U. S.
Department of Energy (Grant No. DE-FG05-96ER40945) and a grant from 
the North Carolina Supercomputing Center.

\end{document}